\documentclass[twocolumn,showpacs,preprintnumbers,nofootinbib,nobibnotes,amsmath,amssymb,aps,prl]{revtex4-1}
\usepackage{todonotes}
\usepackage{hyperref}
\usepackage{slashed}
\usepackage{epstopdf}
\usepackage{amsfonts}
\usepackage{epsfig} 
\usepackage{graphicx,graphics}

\usepackage{xcolor}

\begin{document} 
\preprint{INR-TH-2018-021}

\title{Affleck--Dine baryogenesis via mass splitting}

\author{Eugeny Babichev$^a$, Dmitry Gorbunov$^{b,c}$, Sabir Ramazanov$^d$}

\affiliation{$^a$Laboratoire de Physique Th\'eorique, CNRS, Univ. Paris-Sud, Universit\'e Paris-Saclay, 91405 Orsay, France,\\ 
$^{b}$Institute for Nuclear Research of the Russian Academy of Sciences, 60th October Anniversary prospect 7a, Moscow 117312, Russia\\ 
$^c$Moscow Institute of Physics and Technology, Institutsky per. 9, Dolgoprudny 141700, Russia\\ 
$^d$CEICO, Institute of Physics of the Czech Academy of Sciences, Na Slovance 2, 182 21 Praha 8, Czechia}

\date{\today}

\begin{abstract}
We introduce a class of non-supersymmetric models explaining baryogenesis {\it a la} Affleck--Dine, which use a decay of two superheavy scalar fields with close masses. These scalars acquire non-zero expectation values during
inflation through linear couplings to a function of an inflaton.
After the inflaton decay, the model possesses approximate
$U(1)$-invariance, explicitly broken by a small mass splitting. This
splitting leads to the baryogenesis in the early Universe. Resulting baryon asymmetry is automatically small for the scalars with the masses about the Grand Unification scale and larger. It is fully determined by the inflaton dynamics
and the Lagrangian parameters, i.e., is independent of initial
pre-inflationary conditions for the scalars. 
As a consequence, baryon perturbations are purely adiabatic. We point out a possible origin of the mass
splitting: masses of scalars degenerate at some large energy scale may
acquire different loop corrections due to the interaction with 
the inflaton. Compared to electroweak baryogenesis and conventional Affleck--Dine scenarios, our mechanism generically leads to the proton decay suppressed by the powers
of the superheavy scalar masses, which makes this scenario potentially testable. 
\end{abstract}


\maketitle

{\it Introduction.} 
The matter-antimatter asymmetry of the Universe is one of the mysteries in cosmology, which strongly hints the existence of physics beyond the Standard Model (SM) of particles. Perhaps the most widely discussed class of baryogenesis models 
involves heavy sterile neutrinos~\cite{Fukugita:1986hr}. Lepton asymmetry generated through their decays gets reprocessed into baryon asymmetry (BA) via the sphaleron jumps at the electroweak phase transition~\cite{Kuzmin:1985mm}. 
Another popular way to tackle the problem was suggested by Affleck and Dine 
in Ref.~\cite{Affleck:1984fy}.

The Affleck--Dine (AD) mechanism introduces a $U(1)$-charged scalar
condensate $\Psi$ in the early Universe. Later on the charge is
converted into observed BA through a decay of the condensate into
quarks, see, e.g., Refs.~\cite{Dolgov:1991fr,Riotto:1999yt} for the
reviews. The natural environment, where the AD mechanism operates, is
provided in the supersymmetric extensions of SM~\cite{Dine:1995kz}. In
this picture, the scalar condensate is formed along the flat
directions of its potential, which are inherent in
supersymmetry. However, non-observation of supersymmetry in the
collider experiments motivates to look for different realizations of
the AD mechanism. Then, in conventional AD scenarios one typically
overproduces baryon isocurvature perturbations~\cite{Enqvist:1998pf,
  Koyama:1998hk} in conflict with the cosmological
data~\cite{Akrami:2018odb}. Furthermore, in conventional AD scenarios
resulting BA is too large, unless the reheating temperature is
low. In the present work we propose a realization
of the AD mechanism, which does not assume the existence of flat
directions and automatically avoids the above mentioned problems.

We consider two scalar fields $\Psi_i$, where $i=1,2$, linearly coupled to some function $F(\phi)$ of the inflaton $\phi$, cf.~Ref.~\cite{Babichev:2018afx},
\begin{equation}
\nonumber 
\propto \Psi_i F(\phi) \; .
\end{equation}
These couplings induce non-zero expectation values $\langle \Psi_i \rangle \propto F(\phi)$. 
We assume that the fields $\Psi_i$ are very heavy, i.e., their masses
$M_i \gtrsim H$, where $H$ is the inflationary Hubble parameter. In
this case the fields relax to the expectation values $\langle \Psi_i \rangle$ 
within a few Hubble times independently of their initial values. As
the inflaton decays after inflation, $F (\phi) \rightarrow 0$, the fields $\Psi_i$ 
start oscillating around zero with the amplitudes set by their expectation values during inflation. If $M_1=M_2$, the system possesses $U(1)$-invariance 
with respect to global rotations of the complex field $\Psi=\Psi_1+i\Psi_2$. We associate this $U(1)$-invariance with the baryon symmetry. Successful 
baryogenesis requires baryon number violation. We achieve it by
assuming a small mass splitting, $|M_1-M_2|/|M_1+M_2| \ll 1$, which slightly breaks $U(1)$-symmetry. The resulting
baryon charge is converted into the SM sector through the decay of the condensate $\Psi$ into quarks.

Note that BA generated in this way is fully independent of the initial
conditions for the fields $\Psi_i$. This is in contrast with the standard AD mechanism, where 
the parameter set includes the initial configuration of the fields
$\Psi_i$. In our version of the model baryon isocurvature perturbations are automatically 
suppressed, as we deal with
superheavy fields, cf. Ref.~\cite{Bettoni:2018utf}. Finally, compared to some realizations of the AD mechanism, in our case BA is naturally 
small, i.e., we do not need to introduce very small coupling constants.

{\it The model.} Consider the following action:
\begin{equation}
\nonumber 
S=\int d^4 x \sqrt{-g} \left[ \frac{1}{2}\left|\partial_{\mu}\Psi \right|^2-\frac{1}{2}M^2 \left|\Psi \right|^2\right]+S'+S'' \; .
\end{equation}
Here $\Psi=\Psi_1+i\Psi_2$ is a complex scalar with the mass $M$. In the absence of terms $S'$ and $S''$ the model 
possesses $U(1)$-symmetry, which we identify with the baryon
invariance. The symmetry breaking necessary for the generation of BA is encoded in terms $S'$ and $S''$ defined as 
\begin{equation}
\label{interaction}
S'=-\int d^4 x \sqrt{-g} \left[ \alpha_1 \Psi_1 +\alpha_2 \Psi_2  \right] F(\phi) \; ,
\end{equation}
and 
\begin{equation}
\nonumber 
S''=-\int d^4 x \sqrt{-g} \cdot V(\Psi, \Psi^{*}) \; ,
\end{equation}
respectively. Here $F(\phi)$ is a function of the inflaton $\phi$. We
assume that the inflaton $\phi$ is a canonical scalar slowly rolling the
slope of its flat potential $U(\phi)$. Note that the mechanism of BA
generation we propose works for fairly arbitrary functions $F(\phi)$. We only assume that $F(\phi) \rightarrow 0$ as $\phi \rightarrow 0$.
 The coupling constants $\alpha_i$, $i=1,2$, which we choose to
 be dimensionless, measure the strength of interactions
 between the fields $\Psi_i$ and the inflaton. These
 couplings explicitly break $U(1)$-invariance of the
 model. Nevertheless, the term $S'$ alone does not lead to 
 production of the baryon charge, as we show below. Its role is to give non-zero expectation 
values for the fields $\Psi_i$, i.e., $\langle \Psi_i \rangle \neq 0$,
which is crucial for the AD mechanism. Baryon charge production is
naturally triggered by the symmetry breaking potential $V(\Psi, \Psi^*)$. Like in the standard AD scenarios it is assumed 
to be small compared to the other terms in the action. Generically, from the effective field theory point of view, 
one expects terms of the form $G_n (\phi) \Psi^n_i$ to be present in the action, where $n \geq 2$, 
and $G_n(\phi)$ are some functions of the inflaton. In what follows, we assume that those terms are negligible compared 
to the linear ones in the fields $\Psi_i$.

For $M \gtrsim H$, where $H$ is the Hubble parameter during inflation, the fields $\Psi_i$ quickly relax to their effective minima, which are offset from zero due to the interaction with the inflaton: 
\begin{equation}
\label{expectationvalues}
\Psi_i=-\frac{\alpha_i F(\phi)}{M^2} \; .
\end{equation}
The requirement that the fields $\Psi_i$ are spectators (they do not affect dynamics during inflation) is fulfilled if 
\begin{equation}
\label{spectator}
\frac{\alpha^2_i F^2(\phi)}{M^2} \ll U(\phi) \,.
\end{equation}
One remark is in order here. Were the fields $\Psi_i$ light, i.e., $M \ll H$, the interaction with the inflaton would not be necessary to generate BA. Indeed, the condensate $\langle \Psi_i \rangle \neq 0$ is formed automatically 
because the fields $\Psi_i$ starting from generically non-zero values are in the slow roll regime during inflation. However, in this case considered in some details in Appendix~A, the evolution of the fields $\Psi_i$ and hence the resulting BA 
strongly depend on their initial conditions 
set prior to inflation. On the contrary, in our scenario the solutions~\eqref{expectationvalues} are the attractors and thus initial conditions for the fields $\Psi_i$ are completely irrelevant. 

Despite the explicit breaking of $U(1)$-invariance by the interaction with the inflaton condensate, no baryon charge is produced during inflation. This immediately follows from Eq.~\eqref{expectationvalues} and the expression for the Noether charge density: 
\begin{equation}
\label{noeth}
Q=\Psi_1 \dot{\Psi}_2-\Psi_2 \dot{\Psi}_1 \; .
\end{equation}
In terms of the amplitude $\lambda$ and the phase $\varphi$ of the complex field, $\Psi=\lambda e^{i\varphi}$, the Noether (baryon) charge density is given by $Q=\lambda^2 \dot{\varphi}$. Hence, 
the fact that $Q=0$ means $\dot{\varphi}=0$, so the relative phase of the fields $\Psi_i$ remains frozen with time, 
$\tan \varphi=\frac{\alpha_2}{\alpha_1}$. Note that including the potential $V(\Psi, \Psi^*)$ does not alter this conclusion, but leads to inessential shifts of the fields $\Psi_i$. The non-zero 
phase $\varphi \neq 0$ ensures CP-violation in the model---one of necessary conditions of baryogenesis.

This behavior is crucially different from that in the conventional AD scenarios, where the phase is 
not fixed by the model parameters, but depends on the initial configuration of the complex field. 
In the standard case, the phase field can acquire large baryon isocurvature
perturbations~\cite{Enqvist:1998pf, Koyama:1998hk} in contradiction with the data~\cite{Akrami:2018odb}. 
This is a rather common outcome of the AD mechanism, albeit not unavoidable. 
In our case, perturbations of the fields $\Psi_i$ are highly adiabatic. Indeed, 
for $M \gtrsim H$, any admixture of isocurvature fluctuations quickly relaxes to zero within a few Hubble times. Thus, 
the CP-violating phase $\varphi$ is homogeneous, i.e., fixed at the value $\varphi=\arctan \frac{\Psi_2}{\Psi_1}$, 
modulo adiabatic perturbations at the level ${\cal O}(10^{-5})$.

After inflation, the field $\phi$ decreases, hence $F(\phi) \rightarrow 0$, and the expectation values of the fields $\Psi_i$ relax to 0; the fields $\Psi_i$ start oscillating. 
In this regime the baryon charge is 
produced due to the potential $V(\Psi, \Psi^*)$. In the standard AD scenarios quartic and higher order potentials $V(\Psi, \Psi^*)$ 
are normally used to produce BA. Here we put forward another mechanism of violating 
the baryon symmetry, which occurs through the quadratic potential: 
\begin{equation}
\nonumber  
V(\Psi, \Psi^*)=\frac{\beta M^2}{4} \left(\Psi^2+\Psi^{*2} \right) \, .
\end{equation}
Here $\beta $ is a dimensionless constant describing the splitting
between the masses of the fields $\Psi_i$, 
\begin{equation}
\nonumber 
M^2_1=M^2(1+\beta), \qquad M^2_2=M^2(1-\beta) \; .
\end{equation}
We assume $|\beta| \ll 1$, so that $\beta \approx (M_1-M_2)/M$.

The mechanism of baryon charge generation is as follows. After
inflation, the fields $\Psi_i$ evolve as free heavy scalar fields. They undergo 
rapid oscillations with the amplitudes decreasing with the scale factor $a$: 
\begin{equation}
\label{solutions}
\Psi_{i} \approx -\frac{\alpha_{i} A F_e }{M^2} \cdot \left(\frac{a_e}{a}\right)^{3/2} \cdot \cos \left[M_{i} \cdot (t-t_e) +\theta_c \right] \; .
\end{equation}
Hereafter the subscript $'e'$ stands for the end of inflation; $F_e \equiv F (\phi_e)$. The coefficient $A$ and the phase $\theta_c$\footnote{Both $A$ and $\theta_c$ are spatially homogeneous with a high accuracy, because they are determined by the dynamics of the inflaton, which 
is a homogeneous field modulo tiny perturbations.} account for the effects of post-inflationary decay of the inflaton. They are not important for understanding the qualitative picture of baryogenesis. 
However, the coefficient $A$ affects the amount of BA generated. We concretize it below, when evaluating BA, see also Appendix~B. Substituting the solutions~\eqref{solutions} into Eq.~\eqref{noeth}, it is straightforward to calculate the Noether charge density:
\begin{equation}
\label{ncosc}
Q (t)\approx \alpha_1 \alpha_2 \cdot \frac{A^2 F^2_e}{M^3} \cdot \left(  \frac{a_e}{a} \right)^3 \cdot  \sin \left[ \beta M\cdot (t-t_e) \right] \; .
\end{equation}
Here we omitted the term oscillating with the frequency of order $M$ and kept only the contribution characterized by the reduced frequency $\omega =\beta M$. 
On the time scales $\tau \simeq \Gamma^{-1}$, where $\Gamma$ is the 
scalar field decay rate into quarks, the former undergoes multiple oscillations, and its contribution to BA is washed out. This is 
not necessarily the case for the term written in Eq.~\eqref{ncosc}. If the mass splitting is small relative to the decay rate $\Gamma$, i.e., 
\begin{equation}
\label{hierarchy}
\Gamma \gg \beta M \; ,
\end{equation}
this term changes slowly on the time scales $\tau$ and thus sources BA. Provided that the hierarchy~\eqref{hierarchy} holds, 
one can replace the sine in Eq.~\eqref{ncosc} by its argument: 
\begin{equation}
\nonumber
Q (t) \approx \alpha_1 \alpha_2\cdot  \beta \cdot \frac{A^2 F^2_e}{M^2} \cdot \left(  \frac{a_e}{a} \right)^3 \cdot  t \; .
\end{equation}
Here we also replaced $t-t_e$ by $t$ assuming $t \gg t_e$. We see that the baryon charge density linearly grows with time until $t \sim \Gamma^{-1}$, when 
it gets converted into the standard matter--antimatter asymmetry. In
what follows, we do not assume any particular value of $\Gamma$, but  
it must exceed the decay rate of the fields $\Psi_i$ triggered by the interaction~\eqref{interaction}.

{\it Evaluating baryon asymmetry.} To evaluate the Noether charge density $Q(t)$, we should concretize the function $F(\phi)$. We choose it as follows: 
\begin{equation}
\label{choice}
F(\phi)=\frac{1}{M_{Pl}} \cdot T^{\mu}_{\mu} \, .
\end{equation}
Another choice is a power law dependence of $F(\phi)$ considered later. 
Eq.~\eqref{choice} contains the Planck mass $M_{Pl}$ and the trace of the inflaton
energy momentum tensor, $T^{\mu}_{\mu}=-(\partial_{\mu} \phi)^2+4U(\phi)$. 
Note that in the weak coupling regime, $\alpha_i \lesssim 1$, the function~\eqref{choice} satisfies 
the constraint~\eqref{spectator} for any $M \gtrsim H$. The Noether charge density produced with~\eqref{choice} is given by 
\begin{equation}
\label{nc}
Q (t)\simeq \frac{9\alpha_1 \alpha_2 \cdot \beta }{4 \pi^2} \cdot \frac{A^2 \cdot H^4_e \cdot M^2_{Pl}}{M^2} \cdot \left(  \frac{a_e}{a(t)} \right)^3 \cdot t \; .
\end{equation}
We see that the baryon charge is fully described by the inflaton dynamics (encoded in $H_e$ and $a_e$) and parameters of the Lagrangian of the fields $\Psi_i$. 
As it follows, BA $\Delta_B \simeq \frac{Q}{s}$ grows linearly with time during the hot stage. This growth is cut off at the time $t \sim \Gamma^{-1}$ corresponding 
to the decay of the $\Psi$-condensate to quarks. 
Hence, observed BA can be estimated as $Q/s$ at $t \simeq \Gamma^{-1}$: 
\begin{equation}
\label{bagen}
\Delta_B \simeq \frac{50\alpha_1 \alpha_2  }{\pi^4g_*} \cdot \frac{\beta M}{\Gamma} \cdot \frac{A^2 \cdot H^4_e \cdot M^2_{Pl}}{M^3 \cdot T^3_{reh} } \cdot \left(\frac{a_e}{a_{reh}} \right)^3\; .
\end{equation}
We made use of the equilibrium expression for the entropy density: 
\begin{equation}
\label{entestimate}
s=\frac{2\pi^2 g_* }{45} \cdot T^3,
\end{equation} 
where $g_*$ is the number of ultra-relativistic degrees of freedom. Note that 
at the times of interest essentially all the SM species are ultra-relativistic, i.e., $g_* \gtrsim 100$.

To get an idea of the parameter space in the model, let us first assume the immediate conversion of the inflaton energy density into radiation. 
In the limit of an instant preheating, so that the inflaton and hence
$F(\phi)$ drop to zero abruptly as inflation ends, we have $A \simeq 1$. 
In a more generic case discussed later, the coefficient $A$ can be substantially smaller. Setting $a_e=a_{reh}$ we have
\begin{equation}
\label{tempest}
T=T_{reh}= \left(\frac{45}{4\pi^3 g_*} \right)^{1/4} \cdot  (H_e \cdot M_{Pl})^{1/2} \; .
\end{equation} 
Thus we obtain
\begin{equation}
\label{barestimate}
\Delta_B \simeq \frac{10\alpha_1 \alpha_2}{\pi^{7/4} g^{1/4}_{*}} \cdot \frac{\beta M}{\Gamma} \cdot \left(\frac{H_e}{M} \right)^{5/2} \cdot \left(\frac{M_{Pl}}{M} \right)^{1/2} \; .
\end{equation}
For instance, taking $M \simeq M_{Pl}$, from Eq.~\eqref{barestimate} we find that only a tiny amount of BA
 can be produced in that case, unless $\alpha_{i} \gg 1$, which would imply a strongly coupled regime. 
On the contrary, for $M \simeq H_e$ (the lowest possible value), the required amount of BA can be generated in the weak coupling regime. For the high scale inflation with $H_e \simeq 10^{13}$~\mbox{GeV}, the set of parameters $\alpha_{i} \simeq 10^{-5}$ and $\beta M/ \Gamma \simeq 10^{-3}$ would do the job.

The mechanism of BA generation may also occur through renormalizable interactions between the fields $\Psi_i$ and the inflaton. Consider the function $F(\phi)$ of the form 
\begin{equation}
\label{renorm}
F(\phi)=\phi^3 \; .
\end{equation}
The baryon charge generated by the time $t \gtrsim \Gamma^{-1}$ is given by 
\begin{equation}
\nonumber 
Q=\alpha_1 \alpha_2 \cdot A^2 \cdot \frac{\beta M}{\Gamma} \cdot \frac{\phi^6_e}{M^3} \cdot \left(\frac{a_e}{a} \right)^3 \; .
\end{equation}
Hence, BA is given by 
\begin{equation}
\nonumber 
\Delta_B \simeq \frac{45 \alpha_1 \alpha_2}{2\pi^2g_*} \cdot A^2 \cdot \frac{\beta M}{\Gamma} \cdot \frac{\phi^6_e}{M^3 \cdot T^3_{reh}}  \cdot \left(\frac{a_e}{a_{reh}} \right)^3 \; .
\end{equation}
Consistency of our discussion requires that the condition~\eqref{spectator} is obeyed. This sets the upper bound on the coupling constants $\alpha_i$, which cannot be larger than
\begin{equation}
\nonumber 
\alpha_{max} \simeq \frac{M \sqrt{U(\phi_*)}}{\phi^3_{*}} \; .
\end{equation}
Hence, for the super-Planckian fields $\phi$ characteristic for chaotic inflation, we are always in the weak coupling regime, i.e., $\alpha_{max} \ll 1$. 
Here the subscript $'*'$ denotes the moment of time deep in the
inflationary epoch, when $M \sim H$, and the fields $\Psi_i$ start rolling towards 
their effective minima. If $M \gg H$ throughout inflation, then the moment $'*'$ coincides with the beginning of inflation. In the instant preheating approximation, resulting BA can be written as follows 
\begin{equation}
\nonumber 
\Delta_B \simeq \frac{1}{\pi^{3/4} g^{1/4}_*}\cdot \frac{\alpha_1 \alpha_2}{\alpha_{max}^2}  \cdot \frac{\beta M}{\Gamma} \cdot \left(\frac{\phi_{e}}{\phi_{*}} \right)^{6}\cdot \frac{H^2_* \cdot M^{1/2}_{Pl}}{H^{3/2}_e \cdot M}  \; .
\end{equation}
We see that the observed value of BA is easily achieved even for $\alpha_i \simeq \alpha_{max}$ and $\beta M/\Gamma \simeq 1$, if the ratio $\phi_{e}/\phi_*$ is sufficiently small.  

Let us comment on the realistic situation, when the post-inflationary stage is long. The corresponding effects on the evolution of the fields $\Psi_i$ are encoded in the coefficient $A$. We assume that the function $F(\phi)$ is constant during inflation, times $t<t_e$, and drops as a power law at the times $t>t_e$:
\begin{equation}
\label{f}
F(\phi) = F_e \cdot \left(\frac{t_e}{t} \right)^{s} \; .
\end{equation} 
where the power $s$ is model-dependent. For example, for our choice~\eqref{choice} and quadratic inflation, $U(\phi) \propto \phi^2$, one has $s=2$. 
Note that in Eq.~\eqref{f} we neglected oscillations of the inflaton. We have checked that keeping them gives a sub-dominant contribution 
to the final result. In Appendix~B, we show that in the limit of large $Mt_e \simeq M/H_e$, the coefficient $A$ has the following asymptotic behavior: 
\begin{equation}
\label{est}
A \simeq \frac{H_e}{M} \; .
\end{equation}
Notably, this asymptotics is largely independent of the actual value
of $s$ (which can be an integer or fractional number) and the rate of cosmological expansion during the post-inflationary stage. 

Furthermore, assuming the constant equation 
of state $w =\frac{p}{\rho}$ between the end of inflation and reheating, one 
writes 
\begin{equation}
\label{ratio}
\left(\frac{a_e}{a_{reh}} \right)^{3(1+w)}=\frac{\rho_{reh}}{\rho_e} \approx \frac{8\pi^3 g_* T^4_{reh}}{90H^2_eM^2_{Pl}} \; .
\end{equation}
If the Universe enters the matter-dominated stage right after inflation, i.e., $w=0$, then combining Eqs.~\eqref{bagen},~\eqref{est}, and~\eqref{ratio}, we get 
\begin{equation}
\nonumber 
\Delta_B \simeq \frac{40 \alpha_1 \alpha_2}{9\pi} \cdot \frac{\beta M}{\Gamma} \cdot \left(\frac{H_e}{M} \right)^4 \cdot \frac{T_{reh}}{M} \; .
\end{equation} 
Observed BA can be achieved for the set of parameters: $H_e \simeq 10^{13}~\mbox{GeV}$, $M \simeq 10^{15}~\mbox{GeV}$, $T_{reh} \simeq 10^{13}~\mbox{GeV}$ 
and $\alpha_i\sim\beta M/\Gamma\sim 1$. Contrary to the case of conventional Affleck-Dine mechanism, 
one does not need to assume low reheating temperature: BA is naturally small due to the presence of the ratio $(H_e/M)^4$.

{\it Discussions.} We conclude with three comments. First, in the
present work we assume that the mass splitting measured by the constant $\beta$ is an independent model parameter. Let us point out the opportunity that the mass 
difference could be induced through the loop corrections involving a
virtual inflaton. Namely, we have $M_1=M_2=M$ at the scale, where the
effective interaction with the inflaton~\eqref{choice} is induced, that is 
the Planck scale $M_{Pl}$ by our assumption. Disregarding quadratic
divergences, we can estimate 
the loop corrections to the mass splitting as 
\begin{equation}
\label{qc}
\frac{M^2_{1}-M^2_{2}}{M^2} \sim \frac{\alpha^2_1-\alpha_2^2}{4\pi^2}
\,\log \frac{|\Psi |}{M_{Pl}} \; .
\end{equation}
Such corrections would follow, e.g., from the couplings $\Psi_i
(\partial_{\mu} \phi)^2/M_{Pl}$. In a particular model of underlying
theory at $M_{Pl}$ (or other high-energy scale) the relation between
$U(1)$-breaking masses and couplings with the inflaton may be more
complicated, and we do not elaborate more on the subject.

Second, as we have pointed out earlier, baryogenesis may occur through the quartic potential:
\begin{equation}
\label{potential}
V(\Psi, \Psi^*)=\frac{\xi}{4} \left(\Psi^4+{\Psi}^{*4} \right) \; ,
\end{equation}
where $\xi$ is the dimensionless constant. Such symmetry breaking potentials are commonly utilized in the AD scenarios. 
Contrary to the case involving the mass splitting considered above,
the baryogenesis triggered by the potential~\eqref{potential} takes place 
right after inflation, when the fields $\Psi_i$ just start oscillating. Second, the 
production rate of the baryon charge is proportional to $\sin 4\varphi$. The latter vanishes, if the fields $\Psi_i$ are equally coupled to the inflaton, i.e., $\alpha_1=\alpha_2$. 
Hence, non-zero BA is possible only for $\alpha_1 \neq \alpha_2$. One can show that, in the instant preheating approximation, the resulting BA is given by
\begin{equation}
\nonumber 
\Delta_B \simeq \frac{80\xi \alpha_1\alpha_2 (\alpha^2_1-\alpha^2_2)}{\pi^{15/4} g^{1/4}_*} \cdot \left(\frac{H_e}{M} \right)^{11/2} \cdot \left(\frac{M_{Pl}}{M} \right)^{5/2} \; .
\end{equation} 
We have assumed the function $F(\phi)$ as in Eq.~\eqref{choice}. As in the case of the AD baryogenesis through the mass splitting, BA is too small for the Planckian masses $M$, unless the coupling constants are large. 
On the flipside, for $M \ll M_{Pl}$, baryogenesis may occur in the weakly coupled regime. For $H_e \simeq 10^{13}$~\mbox{GeV} and $M \simeq 10^{15}$~\mbox{GeV},  
the possible set of parameters is $\alpha_{i} \simeq 10^{-1}$ and $\xi \simeq 10^{-3}$.

Third, we would like to point out one potentially relevant signature of the mechanism considered: generically it triggers proton $p$ instability. Recall that the perturbative proton decay is forbidded in electroweak baryogenesis and 
conventional Affleck--Dine scenarios, while corresponding non-perturbative effects are exponentially suppressed. In our case, the proton decay is perturbatively possible, it goes as $p \rightarrow \Psi \rightarrow \phi \rightarrow SM~particles$. The process involves virtual fields $\Psi$ and the inflatons, and thus its rate is suppressed by the powers of the mass $M$ and the inflaton mass. That suppression should be strong enough 
to make the proton decay non-observable in current observations. The actual rate depends on the embedding of our mechanism in a concrete particle framework and 
the inflaton interactions with the SM species required to reheat the Universe.

{\it Acknowledgments.} E.B. acknowledges support from PRC CNRS/RFBR (2018--2020) n\textsuperscript{o}1985 ``Gravit\'e modifi\'ee et trous noirs: signatures exp\'erimentales et mod\`eles consistants'',
and from the research program ``Programme national de cosmologie et galaxies'' of the CNRS/INSU, France.~S.R. is supported by the European Regional Development Fund (ESIF/ERDF) and the Czech Ministry of Education, Youth and Sports (M\v SMT) through the Project CoGraDS-CZ.02.1.01/0.0/0.0/15\_003/0000437. 

\section*{Appendix A: Small masses $M \ll H$}

Let us consider the situation, when the fields $\Psi_i$ do not couple to the inflaton. In that situation, they still may have non-zero expectation values $\langle \Psi_i \rangle \neq 0$, if their masses are small relative 
to the Hubble rate during inflation, $M_i \ll H$. The fields $\Psi_i$ evolve in the slow roll regime starting from some initial values $\Psi_{i, 0}$. The slow roll terminates at some moment $t_{osc}$ in the hot epoch, 
when $M_i \simeq H$, and the fields $\Psi_i$ start oscillating about their zero values. Again assuming the small splitting $|\beta| \approx |M_1-M_2|/M\ll 1$ and following the same steps 
as in the main part of the text, one gets for the Noether charge density:
\begin{equation}
\nonumber 
Q (t) \simeq \beta M^2  \Psi_{1,0} \Psi_{2,0} \cdot (t-t_{osc}) \cdot \left(\frac{a_{osc}}{a (t)} \right)^3 \; .
\end{equation}
The linear growth is cut at the times $t-t_{osc} \simeq \Gamma^{-1}$, when the Noether charge $Q(t)$ gets converted to the quark sector. Hence, 
\begin{equation}
\nonumber 
\Delta_B \simeq \frac{\beta M^2  \Psi_{1,0} \Psi_{2,0}}{\Gamma \cdot s (t_{osc})} \; .
\end{equation}
The entropy density is given by Eq.~\eqref{entestimate}, and the temperature is given by Eq.~\eqref{tempest}, where one should set $H \simeq M$. We end up with the following estimate for BA: 
\begin{equation}
\nonumber 
\Delta_B \simeq \frac{4\pi^{1/4} }{g^{1/4}_*}\cdot \frac{\beta M}{\Gamma} \cdot \frac{M \Psi_{1,0} \Psi_{2,0}}{\left(M \cdot M_{Pl} \right)^{3/2}} \; . 
\end{equation}
We see that BA strongly depends on the initial values of the fields
$\Psi_i$. For $\Psi_{i,0}$ of the Planckian order we get   
\begin{equation}
\nonumber 
\Delta_B \simeq \frac{4 \pi^{1/4} }{g^{1/4}_{*}} \cdot \frac{\beta M}{\Gamma} \cdot \left(\frac{M_{Pl}}{M} \right)^{1/2} \; .
\end{equation}
Hence, the required splitting is estimated as 
\begin{equation}
\nonumber 
\beta \simeq 10^{-10} \cdot \frac{\Gamma}{M} \cdot \left(\frac{M}{M_{Pl}} \right)^{1/2} \; .
\end{equation}
This can be used in order to estimate the largest possible mass splitting in this scenario. Substituting $\Gamma \simeq M$ and $M \simeq 10^{-6} M_{Pl}$, 
one gets $\beta_{max} \simeq 10^{-13}$. We conclude that for the Planckian fields $\Psi_i$, generated BA is large, unless  
the mass splitting is tiny.

\section*{Appendix B: Effects of long post-inflationary stage}
Here we estimate the constant $A$ entering Eq.~\eqref{solutions}. Recall that in the approximation of the (almost) instant preheating one has $A \simeq 1$. We show that for any long post-inflationary stage, the coefficient $A$ 
may substantially deviate from unity.

In the presence of the coupling function $F(\phi)$, the general solution
for the fields $\Psi_i$ is given by (we omit $'i'$ below)
\begin{equation}
\label{general}
\begin{split}
\Psi &=-\frac{\alpha \sin (M t) }{a^{3/2}} \int dt \frac{\cos (M t)}{M} F(\phi) a^{3/2} (t) +\\ 
+&\frac{\alpha \cos (M t) }{a^{3/2}} \int dt \frac{\sin (M t)}{M} F(\phi) a^{3/2}(t)  \; .
\end{split}
\end{equation}
We assume that the fields $\Psi$ are at rest by the end of inflation, i.e.,  
\begin{equation}
\label{ic} 
\Psi \left. \right|_{t=t_e} =-\frac{\alpha F_e}{M^2} \qquad  \dot{\Psi} \left. \right|_{t=t_e}=0 \; .
\end{equation}
For $F(\phi)$ we choose the power law behavior as in
Eq.~\eqref{f}. For the sake of concreteness, consider the inflaton
with the quadratic potential $U(\phi) \propto \phi^2$. In
that case we deal with the matter-dominated stage, so that $s=2$, while the scale factor grows as $a(t) \propto t^{2/3}$. 
The generalization to arbitrary $s$ and different types of cosmological expansion is straightforward, and we comment on it below.  

It is convenient to introduce the dimensionless variable
$\xi=M(t-t_e)$. Substituting Eq.~\eqref{f} with $s=2$ into
Eq.~\eqref{general}, and using $a(t) \propto t^{2/3}$, one writes down the solution for the fields $\Psi$, which respects initial conditions~\eqref{ic}:
\begin{equation}
\label{expression}
\begin{split}
\Psi =& \frac{\alpha F_e \cos \xi}{M^2} \cdot  \frac{M t_e}{M t_e +\xi} \cdot \Bigl[ \int^{\xi}_0 d \xi' \frac{M  t_e \sin \xi' }{M t_e+\xi'}-1 \Bigr]\\ 
&-\frac{\alpha F_e \sin \xi}{M^2} \cdot \frac{M t_e}{M t_e +\xi} \Bigl[\int^{\xi}_0 d \xi' \frac{M t_e\cos \xi' }{M t_e+\xi'}+\frac{1}{Mt_e} \Bigr]\; .
\end{split}
\end{equation}
We are  interested in the late time behavior of the fields $\Psi$, i.e., $\xi \gg 1$. The integrals entering the expression above have nice converging properties, so that 
one can replace the upper limits of integration by infinity. In the limit of large $Mt_e$ the integrals of interest have the following asymptotics~\cite{digital}: 
\begin{equation}
\label{as1} 
\int^{\infty}_0 d \xi' \frac{M t_e \sin \xi' }{M t_e+\xi'}=1-\frac{2}{(Mt_e)^2}+{\cal O} \left[\frac{1}{(M t_e)^4} \right] \; ,
\end{equation}
and 
\begin{equation}
\label{as2}
\int^{\infty}_0 d \xi' \frac{M t_e\cos \xi' }{M t_e+\xi'}=\frac{1}{Mt_e}+{\cal O} \left[\frac{1}{(M t_e)^3} \right] \; .
\end{equation}
Neglecting the terms suppressed by $1/(Mt_e)^2$, we write the solution for the fields $\Psi$ as follows: 
\begin{equation}
\nonumber 
\Psi \approx -\frac{2\alpha F_e \sin \xi}{(Mt_e) \cdot M^2} \cdot \frac{Mt_e}{Mt_e +\xi} \; .
\end{equation} 
Comparing the latter with Eq.~\eqref{solutions} we justify our estimate~\eqref{est}. 
We have checked numerically that the same asymptotic behavior as in Eqs.~\eqref{as1} and~\eqref{as2} holds for 
fairly wide range of $s$ in Eq.~\eqref{f}.


\begin{thebibliography}{99}

\bibitem{Fukugita:1986hr}
  M.~Fukugita and T.~Yanagida,
  Phys.\ Lett.\ B {\bf 174} (1986) 45.
  doi:10.1016/0370-2693(86)91126-3

\bibitem{Kuzmin:1985mm}
  V.~A.~Kuzmin, V.~A.~Rubakov and M.~E.~Shaposhnikov,
  Phys.\ Lett.\  {\bf 155B} (1985) 36.
  doi:10.1016/0370-2693(85)91028-7

\bibitem{Affleck:1984fy}
  I.~Affleck and M.~Dine,
  Nucl.\ Phys.\ B {\bf 249} (1985) 361.

\bibitem{Dolgov:1991fr}
  A.~D.~Dolgov,
  Phys.\ Rept.\  {\bf 222} (1992) 309.

\bibitem{Riotto:1999yt}
  A.~Riotto and M.~Trodden,
  Ann.\ Rev.\ Nucl.\ Part.\ Sci.\  {\bf 49} (1999) 35;
  [hep-ph/9901362].

\bibitem{Dine:1995kz}
  M.~Dine, L.~Randall and S.~D.~Thomas,
  Nucl.\ Phys.\ B {\bf 458} (1996) 291;
  [hep-ph/9507453].





\bibitem{Enqvist:1998pf}
  K.~Enqvist and J.~McDonald,
  Phys.\ Rev.\ Lett.\  {\bf 83} (1999) 2510;
  [hep-ph/9811412].

\bibitem{Koyama:1998hk}
  K.~Koyama and J.~Soda,
  Phys.\ Rev.\ Lett.\  {\bf 82} (1999) 2632;
  [astro-ph/9810006].

\bibitem{Akrami:2018odb}
  Y.~Akrami {\it et al.} [Planck Collaboration],
  arXiv:1807.06211 [astro-ph.CO].

\bibitem{Babichev:2018afx}
  E.~Babichev, D.~Gorbunov and S.~Ramazanov,
  Phys.\ Rev.\ D {\bf 97} (2018) no.12, 123543;
  [arXiv:1805.05904 [astro-ph.CO]].

\bibitem{Bettoni:2018utf}
  D.~Bettoni and J.~Rubio,
  Phys.\ Lett.\ B {\bf 784} (2018) 122;
  [arXiv:1805.02669 [astro-ph.CO]].

\bibitem{digital}
https://dlmf.nist.gov/6.12

\end{thebibliography}
\end{document}